\documentclass[12pt]{article}
\usepackage{graphicx}

\def\Re{{\cal R \mskip-4mu \lower.1ex \hbox{\it e}\,}}
\def\Im{{\cal I \mskip-5mu \lower.1ex \hbox{\it m}\,}}
\def\ie{{\it i.e.}}
\def\eg{{\it e.g.}}

\def\sub#1{_{\lower.25ex\hbox{$\scriptstyle#1$}}}
\def\tev{\,{\ifmmode\mathrm {TeV}\else TeV\fi}}
\def\gev{\,{\ifmmode\mathrm {GeV}\else GeV\fi}}
\def\mev{\,{\ifmmode\mathrm {MeV}\else MeV\fi}}
\def\mpl{\ifmmode M_{pl}\else $M_{pl}$\fi}
\def\mpl{\ifmmode \overline M_{Pl}\else $\bar M_{Pl}$\fi}
\def\to{\rightarrow}

\def\subw{_{\rm w}}
\def\mh{\ifmmode m\sbl H \else $m\sbl H$\fi}
\def\mch{\ifmmode m_{H^\pm} \else $m_{H^\pm}$\fi}
\def\mt{\ifmmode m_t\else $m_t$\fi}
\def\mc{\ifmmode m_c\else $m_c$\fi}
\def\mz{\ifmmode M_Z\else $M_Z$\fi}
\def\mw{\ifmmode M_W\else $M_W$\fi}
\def\mws{\ifmmode M_W^2 \else $M_W^2$\fi}
\def\mhs{\ifmmode m_H^2 \else $m_H^2$\fi}   
\def\mzs{\ifmmode M_Z^2 \else $M_Z^2$\fi}
\def\mts{\ifmmode m_t^2 \else $m_t^2$\fi}
\def\mcs{\ifmmode m_c^2 \else $m_c^2$\fi}
\def\mchs{\ifmmode m_{H^\pm}^2 \else $m_{H^\pm}^2$\fi}
\def\ztwo{\ifmmode Z_2\else $Z_2$\fi}
\def\zone{\ifmmode Z_1\else $Z_1$\fi}
\def\mtwo{\ifmmode M_2\else $M_2$\fi}
\def\mone{\ifmmode M_1\else $M_1$\fi}
\def\tb{\ifmmode \tan\beta \else $\tan\beta$\fi}
\def\xw{\ifmmode x\subw\else $x\subw$\fi}
\def\ch{\ifmmode H^\pm \else $H^\pm$\fi}
\def\lum{\ifmmode {\cal L}\else ${\cal L}$\fi}
\def\inpb{\,{\ifmmode {\mathrm {pb}}^{-1}\else ${\mathrm {pb}}^{-1}$\fi}}
\def\infb{\,{\ifmmode {\mathrm {fb}}^{-1}\else ${\mathrm {fb}}^{-1}$\fi}}
\def\epem{\ifmmode e^+e^-\else $e^+e^-$\fi}
\def\ppb{\ifmmode \bar pp\else $\bar pp$\fi}
\def\bsg{\ifmmode B\to X_s\gamma\else $B\to X_s\gamma$\fi}
\def\bsll{\ifmmode B\to X_s\ell^+\ell^-\else $B\to X_s\ell^+\ell^-$\fi}
\def\bstt{\ifmmode B\to X_s\tau^+\tau^-\else $B\to X_s\tau^+\tau^-$\fi}
\def\lamt{\ifmmode \tilde\lambda\else $\tilde\lambda$\fi}
\def\shat{\ifmmode \hat s\else $\hat s$\fi}
\def\that{\ifmmode \hat t\else $\hat t$\fi}
\def\uhat{\ifmmode \hat u\else $\hat u$\fi}

\newskip\zatskip \zatskip=0pt plus0pt minus0pt
\def\matth{\mathsurround=0pt}
\def\lsim{\mathrel{\mathpalette\atversim<}}
\def\gsim{\mathrel{\mathpalette\atversim>}}
\def\atversim#1#2{\lower0.7ex\vbox{\baselineskip\zatskip\lineskip\zatskip
  \lineskiplimit 0pt\ialign{$\matth#1\hfil##\hfil$\crcr#2\crcr\sim\crcr}}}

\def\grtsim{\,\,\rlap{\raise 3pt\hbox{$>$}}{\lower 3pt\hbox{$\sim$}}\,\,}
\def\lsim{\,\,\rlap{\raise 3pt\hbox{$<$}}{\lower 3pt\hbox{$\sim$}}\,\,}

\renewcommand{\thefootnote}{\fnsymbol{footnote}}

\begin{document} \begin{titlepage}
\rightline{\vbox{\halign{&#\hfil\cr
%&DRAFT\cr
&SLAC-PUB-10856\cr
&December 2004\cr}}}
\begin{center}
\thispagestyle{empty} \flushbottom \centerline{ {\Large\bf Warped 
Phenomenology of Higher-Derivative Gravity 
\footnote{Work supported in part
by the Department of Energy, Contract DE-AC02-76SF00515}
\footnote{e-mails:
$^a$rizzo@slac.stanford.edu}}}
\medskip
\end{center}

\centerline{Thomas G. Rizzo$^{a}$}
\vspace{8pt} 
\centerline{\it Stanford Linear
Accelerator Center, 2575 Sand Hill Rd., Menlo Park, CA, 94025}

\vspace*{0.3cm}

\begin{abstract}
We examine the phenomenological implications at colliders for the existence 
of higher-derivative gravity terms as extensions to the Randall-Sundrum model. 
Such terms are expected to arise on rather general grounds, \eg, from 
string theory. In 5-d, if we demand that the theory be unitary and ghost 
free, these new contributions to the bulk action are uniquely of the 
Gauss-Bonnet form. We demonstrate that the usual expectations for the 
production cross section and detailed properties of graviton Kaluza-Klein 
resonances and TeV-scale black holes can be substantially altered by existence 
of these additional contributions. It is shown that measurements at future 
colliders will be highly sensitive to the presence of such terms. 
\end{abstract}

%\vskip0.45in
%\begin{center}

%\end{center}

\renewcommand{\thefootnote}{\arabic{footnote}} \end{titlepage} 

%
%
%  Happy Holidays! 
%
%
%%%%%%%%%%%%%%%%%%%%%%%%%%%%%%%---- Put text here

\section{Introduction}

The Randall-Sundrum(RS) model{\cite {RS}} provides a geometric solution to the 
hierarchy problem through an exponential warp factor whose magnitude is 
controlled by the separation, $\pi r_c$, of two 3-branes embedded in 5-d 
Anti-deSitter space, $AdS_5$. It has been shown 
that this interbrane distance can be naturally stabilized at a value 
necessary to produced the experimentally observed ratio of the weak and 
Planck scales{\cite {GW}}. In its simplest form, SM matter in the RS model 
is confined to 
one of the 3-branes while gravity is allowed to propagate in the bulk. A very 
generic signature of this kind of scenario is the existence of TeV-scale 
Kaluza-Klein(KK) excitations of the graviton with inverse TeV-scale 
couplings to the SM fields. These states will appear as a series of 
spin-2 resonances in a 
number of processes that should be observable at both hadron and 
$e^+e^-$ colliders which probe the TeV-scale. The  
masses and couplings of these KK graviton states 
will be determined with reasonably high precision by future 
collider measurements. Another possible RS signature is 
the copious production of TeV scale black holes, though this is not a 
unique feature of the RS model{\cite {ADD}}. 

One can easily imagine that this simple RS scenario is incomplete from either 
a top-down or bottom-up perspective. We would generally 
expect some `soft' modifications to 
the details of the picture presented above, hopefully 
without disturbing the nice 
qualitative features of the model. One such extension of the basic RS model is 
the existence of higher curvature terms which might be expected on general 
grounds from string theory{\cite {Zwiebach,Mavromatos}} or other possible 
high-scale completions. Once we open the door to such possibilities the number 
of potential new terms in the action can grow rather rapidly as the number 
of scalars that we can form from products of the curvature and Ricci tensors 
as well as the Ricci scalar are enormous. 
A certain general class of such invariants with very 
interesting properties was first generally described 
by Lovelock{\cite {Lovelock}} and, hence, are termed Lovelock invariants. 
(They are also sometimes referred to, apart from a factor of $\sqrt {-g}$, as 
Euler densities since their volume integrals are related to the Euler 
characteristics.) 
The Lovelock invariants come in fixed order, $n$, which 
we denote as ${\cal L}_n$,  
that describes the number of powers of the curvature tensor, contracted in 
various ways, out of which they are constructed. Given a space of 
dimension $D$ the order of these invariants is constrained: For $D=2n$, the 
Lovelock invariant is a topological one and 
leads to a total derivative in the action whereas 
all higher order invariants, $D\leq 2n-1$, can be shown to vanish identically 
by various curvature tensor index symmetry properties. For $D\geq 2n+1$, 
the ${\cal L}_n$ are dynamical objects that once introduced into the action 
for gravity 
can be shown to lead only to second order equations of motion as is the case 
for ordinary Einstein gravity, \ie, no terms with 
derivatives higher than second will appear in the equations of motion  
due to their presence. Generally, arbitrary 
invariants formed from ever higher powers of the curvature tensor will lead to 
equations of motion of ever higher order, \ie, ever more co-ordinate 
derivatives of the metric tensor and graviton field, \eg, terms with quartic 
derivatives. Such theories will lead to very serious problems with both the  
presence of ghosts as well as with unitarity{\cite {Zwiebach}}. The 
Lovelock invariants are constructed in such a way as to be free of these 
problems making them very special and are found to be 
just the forms taken by the higher order curvature terms generated in 
perturbative
string theory{\cite {Zwiebach,Mavromatos}}. This is just what we may have 
expected if string theories are to avoid these unitarity issues. Indeed, if 
one tracks 
potential ghost generating contributions through the fog of higher derivative 
terms in the Lovelock invariants one sees that they vanish identically. 

In the case of 4-d, apart from numerical factors, ${\cal L}_0=1$ while 
${\cal L}_1=R$, the ordinary Ricci scalar.  The invariant of the next order, 
${\cal L}_2$, can be identified with the Gauss-Bonnet(G-B)  
invariant, $R^2-4R_{\mu\nu}R^{\mu\nu}+
R_{\mu\nu\alpha\beta}R^{\mu\nu\alpha\beta}$, which is a topological term 
as well as a total derivative and can be written {\it in 4-d} as 
$\epsilon^{\mu\nu\rho\sigma}\epsilon_{\alpha\beta\gamma\delta}
R_{\mu\nu}~^{\alpha\beta}R_{\rho\sigma}~^{\gamma\delta}$. All 
higher order invariants vanish. When we go to 5-d, as is the case we will 
discuss below, 
all the ${\cal L}_{n \geq 3}$ still vanish as in 4-d but the G-B invariant 
is no longer a total derivative and its presence will modify the results 
obtained from Einstein gravity in the RS model, altering the equations 
of motion. It is interesting to note 
that if we demand the absence of ghosts, \ie, terms with no more than two 
derivatives of the metric,  
then the addition of the G-B piece to the Einstein term is the 
{\it unique} modification of the 5-d RS bulk action. What is particularly 
amazing is that the addition of the G-B term to the conventional RS model  
still allows for a solution with the same qualitative structure as is 
present in the traditional RS model{\cite {KKL}}. 
In fact, generalizing to  $D$ dimensions, Meissner and 
Olechowski{\cite {Meissner}} have shown that RS-like 
solutions exist with the presence of $(D-2)$-branes even allowing for 
all of the non-zero ${\cal L}_n$ contributions to be present in the 
bulk action. 

Some of the modifications of the RS model due the presence of G-B terms have 
been discussed by other authors (\eg, 
Refs. {\cite {KKL,Nojiri,Meissner,Cho,Cho2,Neupane,Charmousis,Brax}}). 
In this paper we are interested in how the presence of the G-B term alters 
the detailed collider 
phenomenology of the RS model. These modifications may occur in 
many different ways and places. 
One can imagine, \eg, that since the G-B term is of higher order, the triple 
graviton coupling in the RS model{\cite {DR}} may be sensitive to its 
existence. While this is certainly true, it turns out that this is not 
the most immediate or sensitive way to probe for the existence of G-B 
contributions to the action. As we will see below, the  
existence of G-B terms will lead to significant changes in the masses and 
SM matter couplings of the Kaluza-Klein gravitons themselves. 
This variation is described by a 
single new parameter that has a rather restricted range in order to avoid 
tachyons being introduced into the model. In addition, we will demonstrate 
that the anticipated production rate for black holes at colliders can be 
significantly increased by the existence of G-B terms in comparison to the 
usual RS scenario. For simplicity, we will ignore issues having to do with the 
radion{\cite {radion}} in what follows 
as this physics may be sensitive to the stabilization mechanism{\cite {GW}}.
The reader must be careful, however, to insure that no tachyons arise in this 
sector of the theory when exploring the model parameter space. We will also 
restrict our analysis to the case where the SM fields are confined to the 
TeV-brane. Generalization to the case of bulk SM fields is 
straightforward{\cite {DHR}}. 

The outline of the paper is as follows: In Section 2 we will present the basic 
Kaluza-Klein formalism and then determine the masses, couplings and 
wavefunctions for the graviton excitations in the RS model in the presence 
of G-B terms. In Section 3 we will discuss how observations of graviton 
resonances and measurements of their properties at colliders can be used to 
determine the value of the new parameter describing the G-B interactions 
or place stringent upper bound on 
its value if no deviations from the conventional RS scenario are observed. 
Section 4 contains a discussion of black hole production at the LHC and how  
it will be substantially enhanced by any appreciable G-B terms. 
We will also show how the modifications to the usual RS picture can lead to a 
significant alteration in the shape and parameter dependence of the black hole 
subprocess cross section. We then  
summarize and conclude. The Appendix outlines the changes to our basic 
formalism which are necessary 
if graviton brane kinetic terms arising from the usual Einstein action  
are also present.

\section{Kaluza-Klein Formalism}

The essential ansatz of the 5-d RS scenario is the existence of a slice of 
warped, Anti-deSitter space bounded by two `branes' which we assume are fixed 
at the $S^1/Z_2$ orbifold fixed points, 
$y=0,\pi r_c$, termed the Planck and TeV 
branes, respectively{\cite {RS}}. We take the 5-d metric describing this 
setup to be given by the conventional expression 
\begin{equation}
ds^2=e^{-2\sigma} \eta_{\mu\nu}dx^\mu dx^\nu-dy^2\,.
\end{equation}
As usual, due to the $S^1/Z_2$ orbifold symmetry we require $\sigma=
\sigma(|y|)$ and, in keeping with the RS solution, we expect $\sigma=k|y|$ 
with $k$ a dimensionful constant of order the fundamental Planck scale. 
As first shown in Ref.{\cite {KKL}} the inclusion of 
G-B terms does not alter this basic setup. 

The action for the model we consider takes the form
\begin{equation}
S=S_{bulk}+S_{branes}\,,
\end{equation}
where 
\begin{equation}
S_{bulk}=\int d^5x ~\sqrt {-g} \Bigg[{M^3\over {2}} R-\Lambda_b+
{\alpha M\over {2}} \Big(R^2-4R_{AB}R^{AB}+R_{ABCD}R^{ABCD}\Big)\Bigg]\,,
\end{equation}
describes the bulk with $M$ being the 5-d fundamental Planck scale, 
$\Lambda_b$ the bulk cosmological constant and $\alpha$ is a dimensionless 
constant of unknown sign 
which we expect to be of order unity based on naturalness arguments.  
Here we note that the upper case Roman indices $A,B,...$ run over $0-4$ while 
Greek indices will continue to run over $0-3$. 
This bulk action contains not only the usual Ricci scalar, $R$, 
of Einstein gravity but also 
the Gauss-Bonnet quadratic curvature term. Similarly  
\begin{equation}
S_{branes}=\sum_{i=1}^2 \int d^4x~\sqrt {-g_i} \Big ({\cal {L}}_i
-\Lambda_i \Big)\,,
\end{equation}
describes the two branes with $g_i$ being the determinant of the 
induced metric and $\Lambda_i$ the associated brane tensions; 
the ${\cal {L}}_i$ describe possible SM fields on 
the branes. In what follows we will assume as usual that the SM fields are 
all localized on the TeV brane at $y=\pi r_c$. For simplicity we have not 
considered potential contributions arising from  
brane kinetic terms for the graviton generated by the Ricci 
scalars evaluated on the two branes{\cite {DHRbt}}; as we will see in the 
Appendix, such contributions can be 
included in a relatively straightforward manner. G-B like brane terms are 
automatically absent as the G-B invariant is at most a surface term in less 
than 5-d. 

To go further we insert our metric ansatz into the equations of motion arising 
from the above action; we obtain $\Lambda_{TeV}=-\Lambda_{Planck}$ as usual 
together with{\cite {RS}} 
\begin{eqnarray}
3\sigma''\Big(1-{4\alpha\over {M^2}} (\sigma')^2\Big) & =& {\Lambda_{Planck}
\over {M^3}}\Delta(y)\nonumber \\
6(\sigma')^2\Big(1-{2\alpha\over {M^2}}(\sigma')^2\Big) &=& -{\Lambda_b 
\over {M^3}}\,,
\end{eqnarray}
where $'=\partial_y$ and $\Delta(y)$ is the combination 
\begin{equation}
\Delta(y)=\delta(y)-\delta(y-\pi r_c)\,.
\end{equation}
As in the conventional RS model the solution indeed takes the form 
$\sigma=k|y|$ though greater than normal care is required to regularize the 
associated $\delta$-functions. Since 
\begin{equation}\
\sigma''=2k\Delta\,,
\end{equation}
using the set of useful relations{\cite {Meissner}}
\begin{equation}
(\sigma')^n \sigma''= {2\over {n+1}} k^{n+1}\Delta(y)\,,
\end{equation}
one then finds 
\begin{equation}
\Lambda_{Planck}=-\Lambda_{TeV}=6kM^3\Big(1- {4\alpha k^2 \over {3M^2}}
\Big)\,. 
\end{equation}
This differs by a factor of $1/3$ in the $\alpha$-dependent term from 
the work of Kim, Kyae and Lee{\cite {KKL}} 
due to an error in their application of the proper boundary 
conditions. In addition, using the relation $(\sigma')^{2n}=k^{2n}$, 
we obtain the following explicit expression for $k$:
\begin{equation}
k=k_\pm=\Bigg [{M^2\over {2\alpha}}\Bigg(1\pm \Big(1+{4\alpha \Lambda_b
\over {3M^5}}\Big)^{1/2}\Bigg)\Bigg]^{1/2} \,.
\end{equation}
Note that for $\alpha>0$, $\Lambda_b$ may in principle take either sign 
whereas it must be negative if $\alpha \leq 0$; the conventional RS limit 
can be obtained by employing the negative root in the equation above with  
$\Lambda_b$ negative and taking the limit $\alpha \to 0$.  

In order to obtain the equations of motion for the KK excitations we need to 
go beyond this vacuum solution; to this end we consider metric perturbations 
of the form 
\begin{equation}
g_{\mu\nu}=e^{-2\sigma}\eta_{\mu\nu}+h_{\mu\nu}\,,
\end{equation}
and employ the traceless-transverse gauge with $\partial^\mu h_{\mu\nu}=
h^\mu_\mu=0$ as usual followed by the KK decomposition
\begin{equation}
h_{\mu\nu}(x,y)=\sum_n h_{\mu\nu}^{(n)}(x)\chi_n(y)\,.
\end{equation}
such that  the Klein-Gordon equation, 
$\partial_\lambda^2 h_{\mu\nu}^{(n)}=-m_n^2 h_{\mu\nu}^{(n)}$, yields the 
KK masses. The $\chi_n$ wavefunctions are then 
seen to satisfy{\cite {Cho,Neupane}}
\begin{equation}
\Bigg[\partial_y^2-4\bar \alpha \sigma'^2\partial_y^2-8\bar \alpha \sigma'
\sigma''\partial_y-4(\sigma')^2(1-4\bar \alpha \sigma'^2)+2\sigma''(1-
12\bar \alpha \sigma'^2)\Bigg]\chi_n= m_n^2e^{2\sigma}(1-4\bar \alpha 
\sigma'^2+4\bar \alpha \sigma'')\chi_n\,,
\end{equation}
where $\bar \alpha= \alpha/M^2$, as obtained by {\cite {Cho}}. Letting $\chi_n=
e^{-2\sigma}\psi_n$, the left-hand side of the expression above can be 
simplified using the identities in Eq.(8) to produce 
\begin{equation}
\Bigg(1-4 {\alpha\over {M^2}} \sigma'^2\Bigg)(\psi_n''-4\sigma'\psi_n')-
8{\alpha\over {M^2}} \sigma' 
\sigma''\psi_n\,,
\end{equation}
which after some algebra directly leads to the G-B 
generalization{\cite{Charmousis}}  
of Eq.(5) in {\cite {DHR}} obtained long ago: 
\begin{equation}
\partial_y \Bigg[\Bigg(1-4{\alpha\over {M^2}}\sigma'^2\Bigg) 
e^{-4\sigma}\partial_y \psi_n\Bigg]+ 
 m_n^2e^{-2\sigma}\Bigg(1-4{\alpha\over {M^2}}\sigma'^2+4{\alpha \over{M^2}}
\sigma''\Bigg)\psi_n=0\,.
\end{equation}
As noted by other authors{\cite {KKL,Cho,Cho2,Neupane,Brax}}, it is 
straightforward to see that this produces the same   
differential equation in the bulk 
for the KK eigenfunctions as found in the usual RS scenario{\cite {DHR}} and 
thus 
\begin{equation}
\psi_n(y)={e^{2\sigma}\over {N_n}} \zeta_2(m_ne^\sigma/k)\,,
\end{equation}
with $\zeta_q=J_q+\beta_n Y_q$ being a combination of Bessel functions 
of order $q$ and 
where $N_n$ is a normalization factor. Following our earlier 
work{\cite {DHR}} we choose to normalize the KK states 
over the full interval $-\pi r_c \leq y \leq \pi r_c$. 
For a differential equation with this truncated Sturm-Liouville 
form{\cite {Arfken}}  
\begin{equation}
{\cal L}\psi_n=[p(y)\psi_n']'+\lambda_n w(y)\psi_n=0\,,
\end{equation}
the eigenfunctions are to be 
orthonormalized with respect to the weight function $w(y)$ 
which we can read off in our case from Eq.(15) above:
\begin{equation}
w(y)=e^{-2\sigma}\Bigg(1-4\alpha {k^2\over {M^2}}+8\alpha {k\over {M^2}}
\Delta\Bigg)\,, 
\end{equation}
so that 
\begin{equation}
\int_{-\pi r_c}^{\pi r_c} ~dy ~w(y) ~\psi_n(y) \psi_m(y)=\delta_{nm}\,.
\end{equation}
Thus, making use of the orbifold symmetry, we obtain the normalization factor  
\begin{equation}
N_n^2=2\int_0^{\pi r_c} dy e^{-2ky}
\Big(1-4\alpha {k^2\over {M^2}}\Big)\psi_n^2(y)+
8\alpha {k\over {M^2}}\Big(\psi_n(0)^2-\epsilon^2\psi_n(\pi r_c)^2\Big)\,.
\end{equation}
where $\epsilon=e^{-\pi kr_c}$.
Neglecting terms suppressed by powers of $\epsilon$, for the massless 
zero-mode graviton this yields the explicit expression 
\begin{equation}
N_0^2={1\over {k}}\Big(1+4\alpha {k^2\over {M^2}}\Big)={1\over {k}}
\Big(1-4\alpha {k^2\over {M^2}}\Big)(1+2\Omega)\,,
\end{equation}
where we have defined for later purposes the combination 
\begin{equation}
\Omega={4\alpha k^2/M^2\over{1-4\alpha k^2/M^2}}\,.
\end{equation}
Knowing this normalization we can directly relate the 5-d 
parameters $M$ and $k$ 
to the (effective) 4-d Planck mass, $\mpl$:
\begin{equation}
\mpl^2={M^3\over {k}}\Big(1-4\alpha {k^2\over {M^2}}\Big)
(1+2\Omega)\,,
\end{equation}
To obtain the explicit expression for the normalization of 
the massive KK excitations we need to examine the two 
boundary conditions which are now quite different than in the usual RS model.  
These are obtained by the integration of Eq.(15) above around the two branes  
at $y=0$ and $\pi r_c$; from this procedure we arrive at  
\begin{equation}
\psi_n(0^+)~'+\Omega{m_n^2\over {k}}\psi(0^+)=0\,,
\end{equation}
and 
\begin{equation}
\psi_n(\pi r_c^+)~'+\Omega \epsilon^{-2}{m_n^2\over {k}}\psi_n(\pi r_c^+)=0\,,
\end{equation}
where the eigenfunctions and their derivatives are evaluated on the positive 
side of both branes. These boundary conditions then determine both 
the wavefunction coefficients 
$\beta_n$ as well as the KK masses $m_n=x_nk\epsilon$; we obtain  
\begin{equation}
\zeta_1(x_n)+\Omega x_n \zeta_2(x_n)=0\,,
\end{equation}
whose roots determine the $x_n$ and 
\begin{equation}
\beta_n=-{{J_1(x_n\epsilon)+\Omega x_n\epsilon J_2(x_n\epsilon)}\over 
{Y_1(x_n\epsilon)+\Omega x_n\epsilon Y_2(x_n\epsilon)}}\,. 
\end{equation}
Note that under {\it most} circumstances $\beta_n \sim \epsilon^2$ and can be 
safely ignored as in the usual RS case; some care, however, needs to be 
exercised as $\Omega \to -1/2$. 
From these expressions we can straightforwardly obtain an explicit expression 
for the normalization 
factor for the KK excitation wavefunctions:
\begin{equation}
N_{n>0}^2={1\over {k\epsilon^2}}\zeta_2(x_n)^2\Big(1-4\alpha {k^2\over {M^2}}
\Big)(1+2\Omega+\Omega^2x_n^2)\,, 
\end{equation}
where subleading terms in powers of $\epsilon$ have been neglected. 
The usual RS model is seen to correspond 
to the limit $\alpha,\Omega\to 0$ in the expressions above. 

At this point the experienced reader may notice that the 
modifications to the RS 
expressions above due to the new G-B interactions is strikingly similar 
in nature to those produced by the addition of conventional $R$-type kinetic 
terms on both branes{\cite {DHRbt}}. This will allow one to combine the 
contributions of both these new physics sources 
into rather simple expressions as we will see in the Appendix below. 

\begin{figure}[htbp]
\centerline{
\includegraphics[width=8.5cm,angle=90]{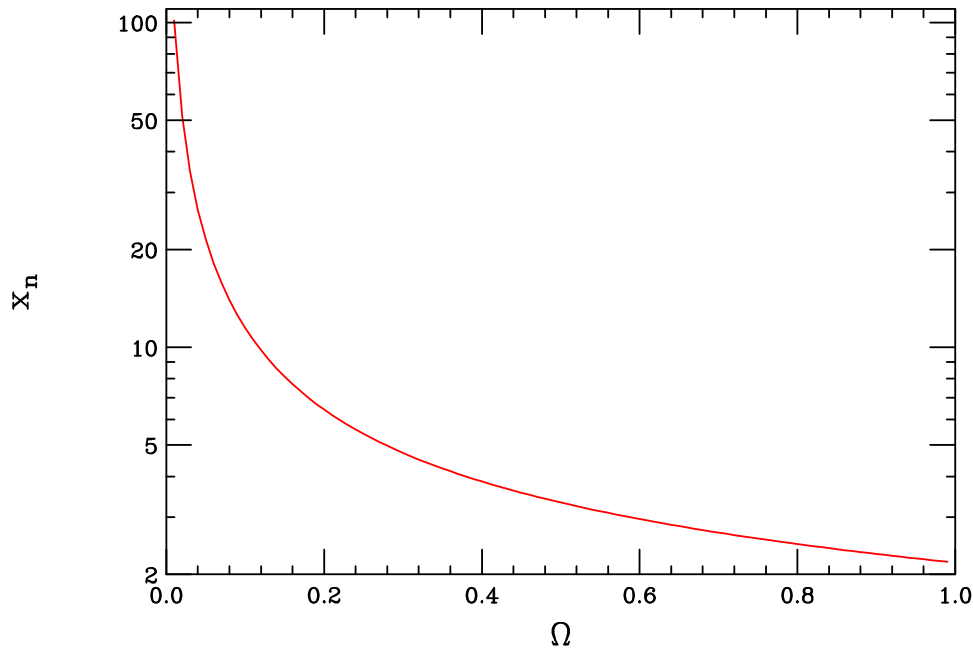}}
\vspace*{0.1cm}
\caption{Imaginary root corresponding to a tachyon in Eq.(26) as a function of 
positive $\Omega$.}
\label{fig1}
\end{figure}

\section{Gravitons at Colliders}

The discussion above tells us that the modifications of the basic RS model 
due to G-B terms can be described by a single parameter, $\Omega$. To go 
further and see this in detail 
we must understand the potential range of this parameter. Since we 
expect $\alpha$ to be of order unity as well as $k\sim M$, one should 
expect that $\Omega$ is also of order unity. The arguments given in 
Ref.{\cite {Charmousis,Brax}} suggest that 
$\Omega$ cannot be positive as this leads to a 
tachyon in the KK spectrum. To verify this claim we have examined Eq.(26) in 
detail when $\Omega>0$ for imaginary roots of the form $z_n=\pm ix_n$ with the 
results shown in Fig.~\ref{fig1}. Here we see that indeed a pair of tachyonic 
roots do exist for $\Omega >0$ which move off to infinity as we approach the 
RS limit $\Omega\to 0$. We thus agree that solutions with 
positive values of $\Omega$ are indeed  
excluded. A similar search for tachyonic roots for $\Omega <0$ was 
unsuccessful  
so we conclude that $\Omega \leq 0$ to which we now restrict ourselves; this 
implies that the parameter $\alpha$ is also $\leq 0$.  
Eq.(21) provides the normalization for the zero-mode graviton wavefunction 
which must be positive definite in order to avoid ghosts. We notice that this 
requires $4\alpha k^2/M^2 >-1$ which then implies, via the definition Eq.(22) 
and the negative $\Omega$ constraint, that $\Omega$ is confined to the range  
\begin{equation}
-{1\over {2}}< \Omega \leq 0\,. 
\end{equation}
A short analysis shows that the consideration of the the normalization factor 
for the graviton KK 
excitations will not improve upon this bound. Thus $\Omega$ is constrained 
to be in a rather narrow range which significantly increases the 
predictability of the model. Given the definition of $\Omega$, Eq.(29) 
implies that 
\begin{equation}
-\alpha {k^2\over {M^2}} \leq {1\over {4}}\,. 
\end{equation}

\begin{figure}[htbp]
\centerline{
\includegraphics[width=8.5cm,angle=90]{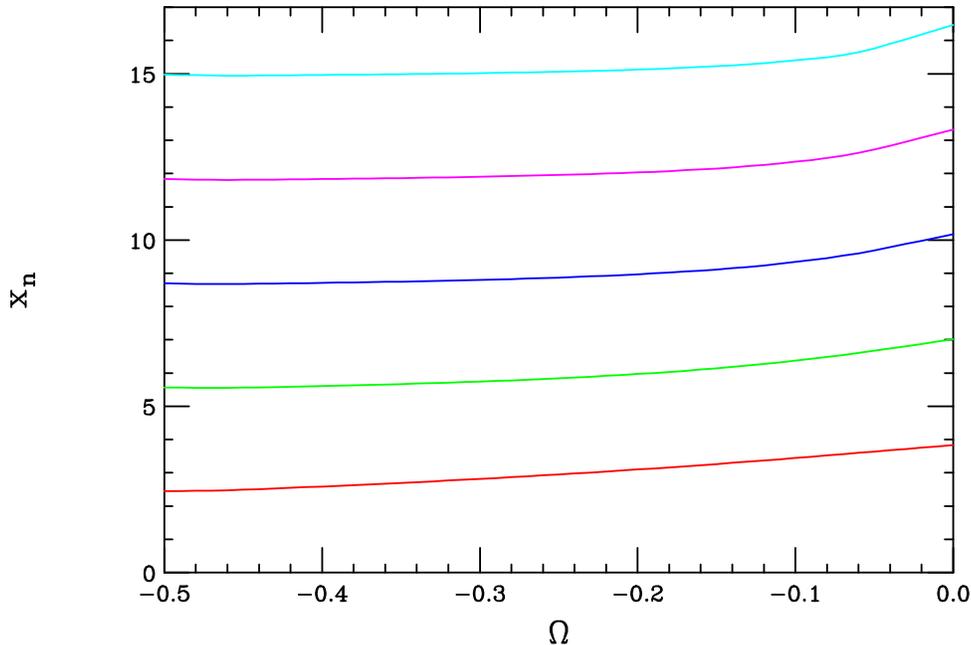}}
\vspace*{0.1cm}
\caption{Behaviour of the first five roots corresponding to the 
graviton KK masses $x_n=m_n/k\epsilon$ as functions of the parameter $\Omega$.}
\label{fig2}
\end{figure}

To begin a phenomenological analysis the 
first question to address is: how does the KK graviton spectrum shift as we 
turn on a non-zero $\Omega$? To this end we extract the first few 
roots, $x_n$, 
provided by Eq.(26) and follow their $\Omega$-dependence; this is shown in 
Fig.~\ref{fig2}. Here we see that the overall $\Omega$ dependence of the roots 
is rather weak across the rather restricted allowed range for $\Omega$. 
The values 
of the roots are seen to decrease slightly as $\Omega$ moves away from 0. 
These small variations in the root values will, however, be of interest to us 
below and will provide an important test of the model. 

\begin{figure}[htbp]
\centerline{
\includegraphics[width=8.5cm,angle=90]{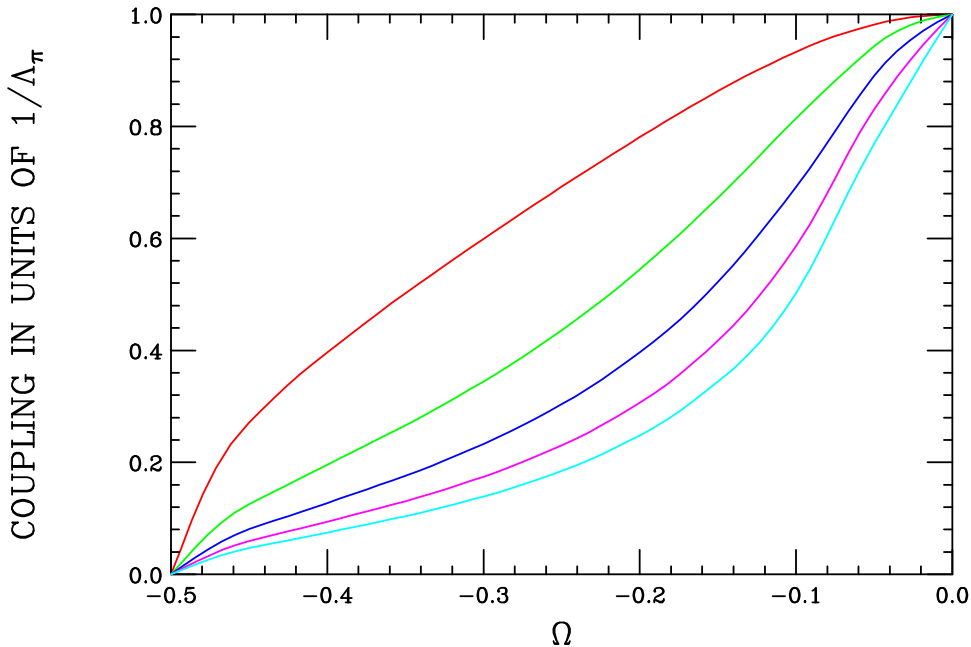}}
\vspace*{0.1cm}
\caption{Coupling strengths of the first KK graviton states, from top to 
bottom, in units of 
$1/\Lambda_\pi$ as functions of the parameter $\Omega$. Note that in the RS 
limit all states have the same coupling.}
\label{fig3}
\end{figure}

Next, we find that the interaction of the KK graviton excitations 
with the SM fields on the TeV brane is given by 
\begin{equation}
{\cal L}={1\over {\Lambda_\pi}} \sum_n  \Bigg[{{1+2\Omega}\over {1+2\Omega
+\Omega^2x_n^2}}\Bigg]^{1/2} h^{\mu\nu}_n T_{\mu\nu}\,,
\end{equation}
where as usual we define $\Lambda_\pi=\mpl\epsilon${\cite {DHR}}. 
As in the case of graviton brane 
kinetic terms, but {\it unlike} in the usual RS model, the couplings of the KK 
graviton tower states have become level-dependent. In particular, 
the KK states 
will always become more weakly coupled as we go up the KK tower. Furthermore,  
we see that the couplings of all the KK states {\it vanish} as  
$\Omega \to -0.5$. (It is interesting to note that a very similar rescaling 
of the conventional RS result holds for the case when SM fields are put in 
bulk.) Fig.~\ref{fig3} shows the rapid decrease in the 
couplings of the first five KK states as $\Omega$ decreases from 0. The 
couplings are observed to vanish more quickly as one goes farther up the KK 
tower as expected. 

In addition to specifying $\Omega$ and the mass of the first KK excitation, 
$m_1$, to completely determine the phenomenology of this model we need to know 
the ratio $c=k/\mpl$ which enters into the expression for the KK widths. In 
the usual RS model this ratio is bounded by the requirement that we don't 
want potential quantum (\ie, higher curvature) corrections to dominate over 
those from classical Einstein gravity. This requirement is traditionally 
stated as $|R|<M^2$, which implies that $k^2/M^2 \leq 1/20$ for a 5-d 
Anti-deSitter space; combining 
with Eq.(23) in the $\alpha \to 0$ limit then implies 
$k/\mpl \lsim 1/10$ which is the usual result. Under the present 
circumstances there may be (at least) three 
objections to this argument. First, 
since we are already including G-B terms there is no longer any reason to 
demand small curvature and thus $k^2/M^2$ may be larger than $1/20$. 
A second objection is that if we do truly want small higher-curvature terms 
for some reason  the traditional constraint on $k^2/M^2$ may be too weak. 
To see this we note that for the RS vacuum solution we know that 
$R=-20k^2$ (hence, the limit above), $R_{AB}R^{AB}=80k^4$ and 
$R_{ABCD}R^{ABCD}=40k^4$ which implies that the G-B invariant has the value 
$120k^4$. These large numerical coefficients may imply that a stronger 
constraint on $k^2/M^2$ might be necessary. A third  
observation is that even if $k^2/M^2 <1/20$ holds, for finite $\alpha$ 
from Eqs.(21) and (23) we find that 
\begin{equation}
{k^2\over {\mpl^2}}={k^3\over {M^3}}\Bigg(1+
4\alpha {k^2\over {M^2}}\Bigg)^{-1}={k^3\over {M^3}}{{(1+\Omega)}
\over {(1+2\Omega)}}\,, 
\end{equation}
so that, given the bound on negative $\Omega$ above, the ratio $k/\mpl$ can 
become {\it arbitrarily} large as $\Omega$ approaches $-0.5$. While we will 
restrict ourselves to the conventional RS range of $k/\mpl$ in what follows 
for purposes of comparison, we remind the reader that much larger values 
of this ratio may now be possible. (There are, however, other 
reasons to believe 
that this ratio may remain as constrained as in the conventional RS model as 
discussed in Ref.{\cite {DHR}}).  Given a values of $\Omega$ and 
the ratio $c=k/\mpl$, we note that 
the parameter $\alpha$ can now be uniquely determined:
\begin{equation}
\alpha={1\over {4c^{4/3}}} {\Omega\over {(1+2\Omega)^{2/3}(1+\Omega)^{1/3}}}
\,.
\end{equation}

\begin{figure}[htbp]
\centerline{
\includegraphics[width=8.5cm,angle=90]{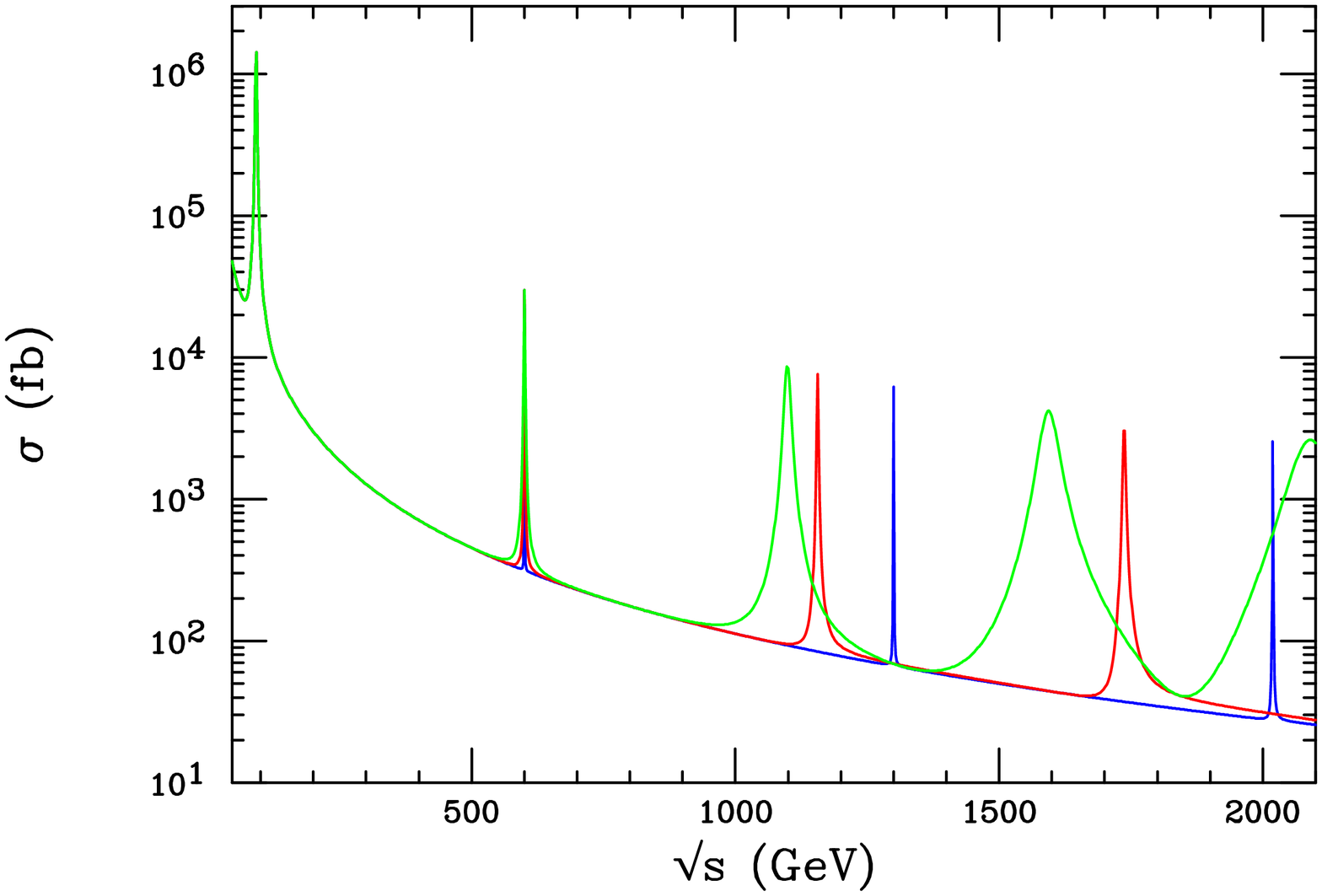}}
\vspace{0.4cm}
\centerline{
\includegraphics[width=8.5cm,angle=90]{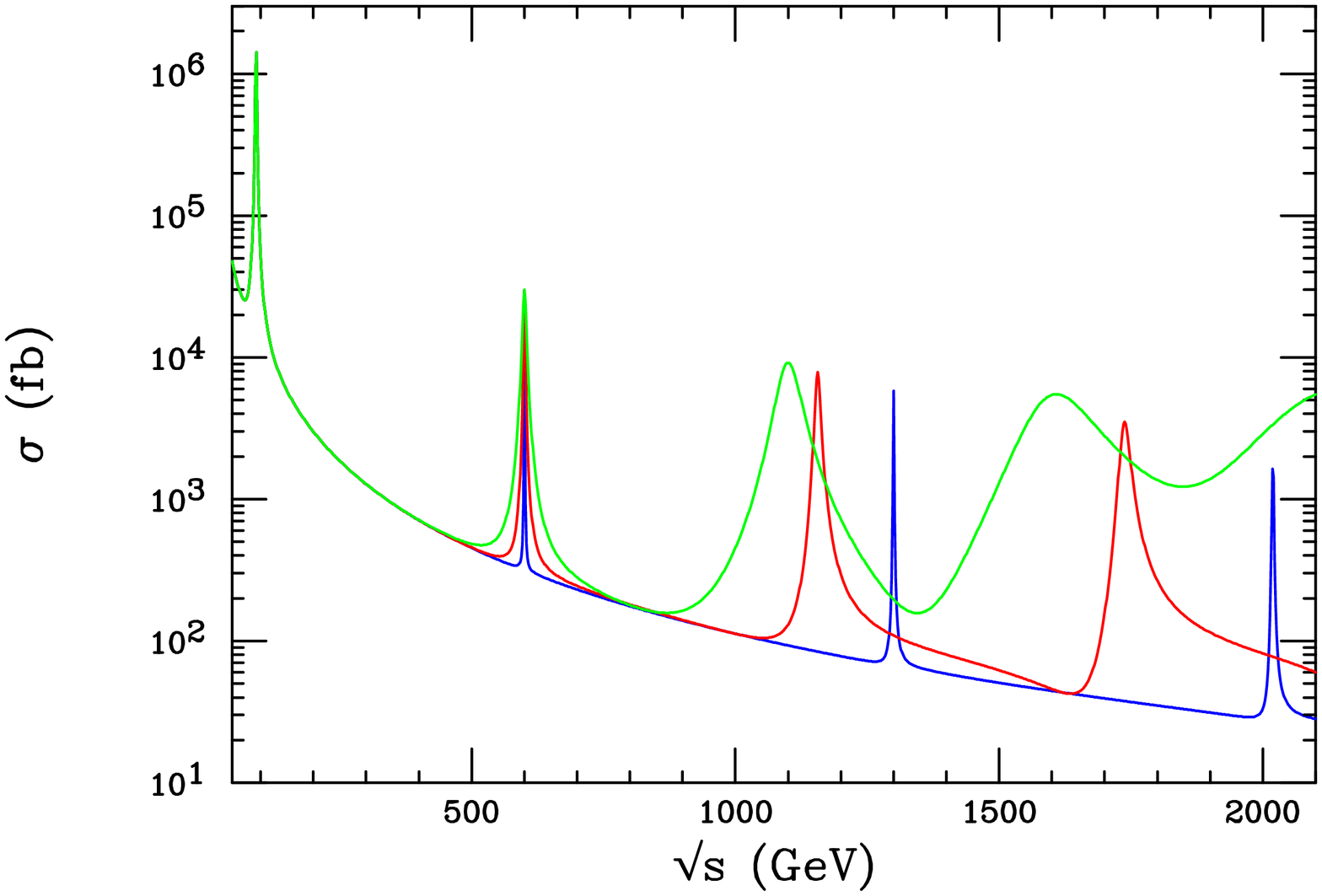}}
\vspace*{0.1cm}
\caption{Cross section for $e^+e^- \to \mu^+\mu^-$ assuming $m_1=600$ GeV 
and $k/\mpl=0.05$(top) or 0.1(bottom). The usual RS model prediction with 
$\Omega=0$ is shown in green while the corresponding results in the present 
model with $\Omega=-0.2$(red) and $-0.4$(blue) are also displayed.}
\label{fig4}
\end{figure}

\begin{figure}[htbp]
\centerline{
\includegraphics[width=8.5cm,angle=90]{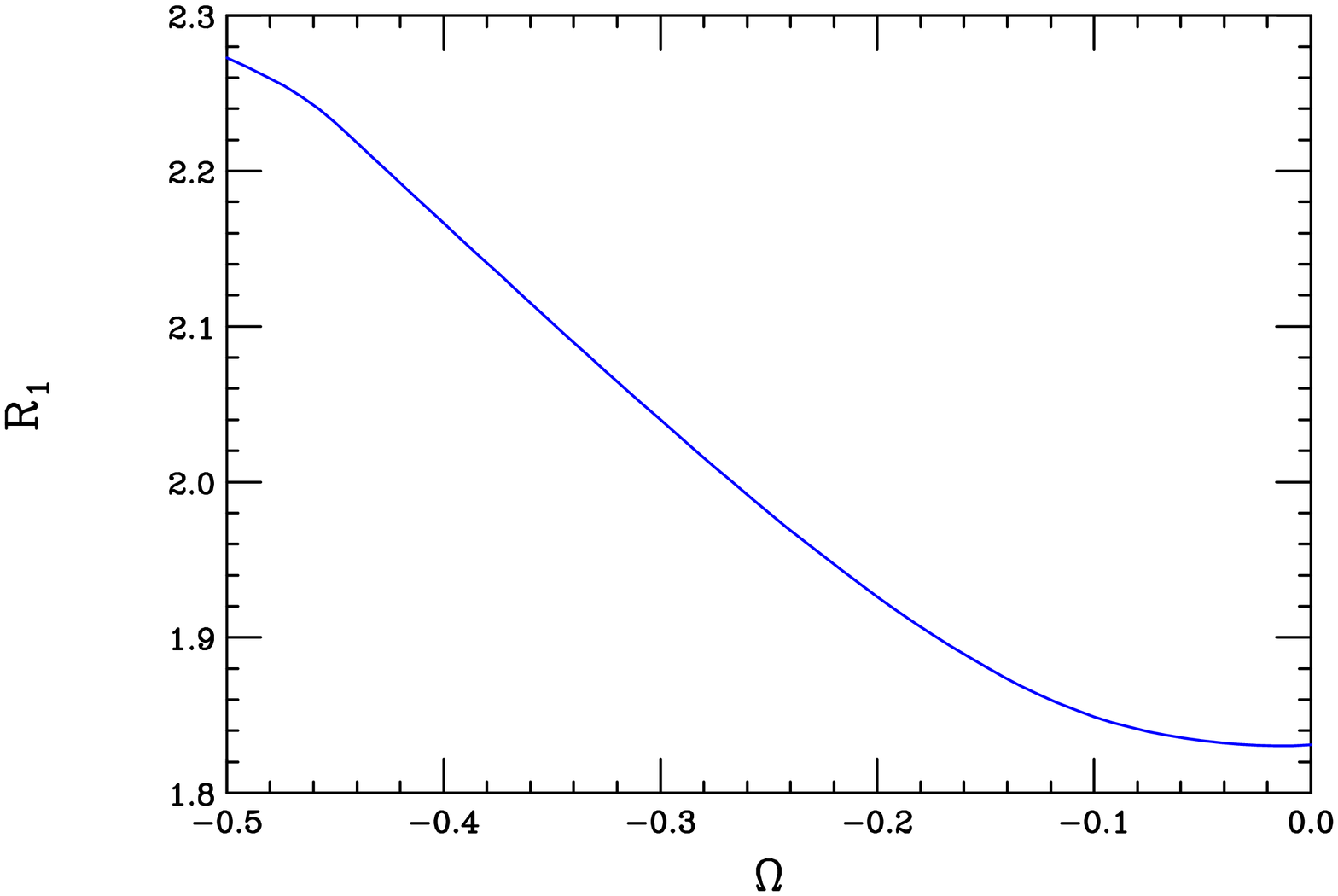}}
\vspace{0.4cm}
\centerline{
\includegraphics[width=8.5cm,angle=90]{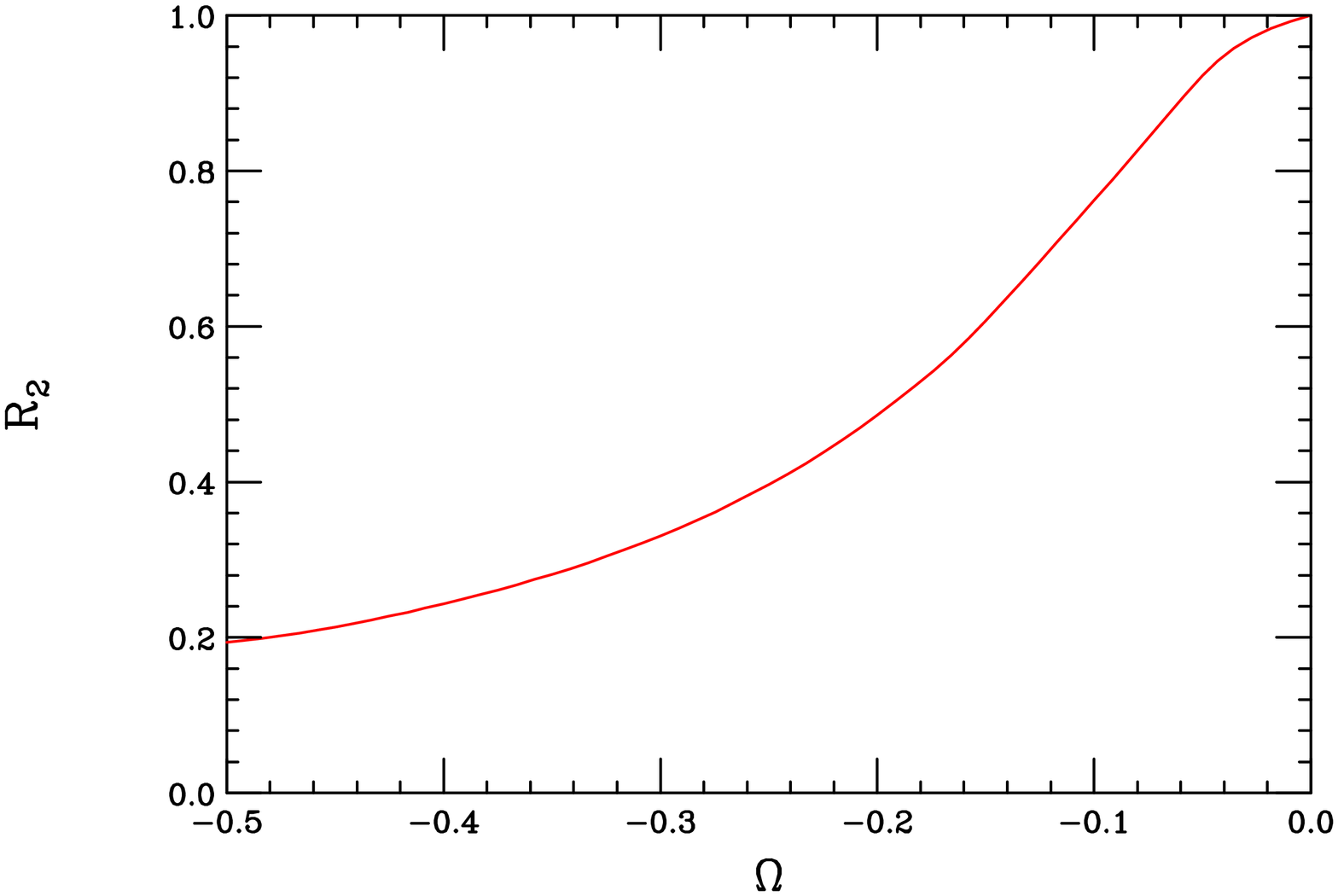}}
\vspace*{0.1cm}
\caption{The ratios $R_{1,2}$ described in the text as functions of $\Omega$.}
\label{fig5}
\end{figure}

With these preliminaries we can now examine how these KK graviton states 
would appear at a collider in comparison to the expectations of the RS 
model. To this end we examine the process $e^+e^- \to \mu^+\mu^-$ in the 
energy range accessible to the International Linear Collider(ILC) and its 
potential upgrades. A similar analysis can, of course, be performed at the 
LHC. To be specific for purposes of demonstration we take $m_1=600$ GeV 
and $k/\mpl=0.05,0.1$;  
the results of these calculations are displayed in Fig.~\ref{fig4} where we 
show the cases $\Omega=-0.2,-0.4$ in comparison to the standard RS model 
predictions with $\Omega=0$. Several things are immediately apparent: 
First, as the magnitude of $\Omega$ increases, the KK states become 
increasingly narrow and their mass splitting is also seen to increase 
significantly. Second, 
with $m_1$ held fixed, the observation of a single graviton resonance 
and a determination of its properties would be insufficient to tell us 
whether or not G-B terms are present in the action. The reason for this is 
clear: if we just observe a single state we can attribute its width 
solely to the value of $\Lambda_\pi$. We see for example that the width of the 
first KK is decreased as $\Omega$ decreases from zero by the square of the 
factor appearing in Fig~\ref{fig3}. The only thing we can extract from the 
first KK width is the value of an {\it effective} $\Lambda_\pi$ which tells 
us nothing about the true value of $\Lambda_\pi$ or $\Omega$. To obtain the 
values of these quantities we need to also find the second KK excitation. 
Fig.~\ref{fig4} shows that the mass of the second KK increases relative to the 
first as the magnitude of $\Omega$ increases; it also gets much more narrow. 
These observations 
provide our best handles on $\Omega$. If we measure the ratio of the 
first two KK masses, $R_1=m_2/m_1$, this will directly tell us the value of 
$\Omega$ as shown in Fig.~\ref{fig5}.

We can get a second handle on $\Omega$ as well as $\Lambda_\pi$ (and thus 
$k/\mpl$ through the relation $m_n=x_n (k/\mpl) \Lambda_\pi$) by comparing the 
decay widths of the first two KK states 
after removing `phase space' factors and knowing  
their respective masses. Since kinematically 
$\Gamma_n \sim m_n^3$, the scaled ratio 
$R_2=\Gamma_2 m_1^3/\Gamma_1 m_2^3$ would be unity in the usual RS model if 
we neglect the masses of the final state particles. This ratio depends 
rather strongly on the value of $\Omega$ when G-B terms are present as is 
shown in Fig.~\ref{fig5}. Recall that as $\Omega$ approaches $-0.5$ the 
couplings of the KK states to SM matter on the TeV brane 
vanishes so that they will no longer 
be produced at colliders. Of course, the {\it ratio} of the couplings of two 
different KK states remains finite. 

\begin{figure}[htbp]
\centerline{
\includegraphics[width=8.5cm,angle=90]{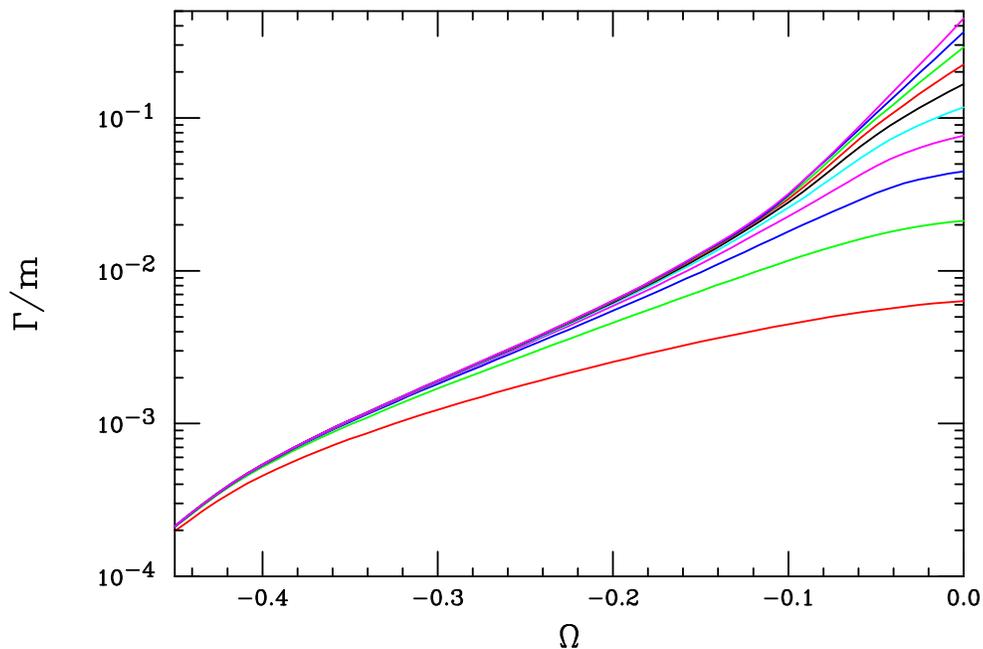}}
\vspace*{0.1cm}
\caption{The ratio $\Gamma/m$ as a function of $\Omega$ for the first ten KK 
states. The KK number goes up as we go from the bottom to the top of the 
figure. $k/\mpl=0.05$ has been chosen for purposes of demonstration.}
\label{fig5p}
\end{figure}

For fixed $\Omega$ and $k/\mpl$ it is interesting to examine the widths 
of the graviton KK resonances as we go farther up the tower. In the usual RS 
scenario the widths grow quite rapidly $\Gamma_n\sim m_n^3$ and eventually the 
resonance structure is lost as can be seen in the lower panel of 
Fig.~\ref{fig5}. Once G-B terms are present, however, the coupling 
decreases as $m_n$ increases and eventually $\Gamma_n \sim m_n$. This means 
that for appreciable values of $\Omega$ the KK states are all becoming rather 
narrow as we go further and further 
up the tower. This can be seen more explicitly by examining the 
ratio $\Gamma_n/m_n$ as a function of $\Omega$ for the different KK tower 
states as is shown in Fig.~\ref{fig5p}. Here we see that for large $\Omega$ 
the KK states remain narrow and have almost the same value of $\Gamma/m$ 
independent of $n$. 

One may wonder just how small a value of $\Omega$ might be measurable. It is 
clear from Figs.~\ref{fig4} and ~\ref{fig5} that the ILC will be highly 
sensitive to small values of $\Omega$ through the mass and width ratios 
discussed above. Note in particular that in the region near $\Omega=0$ 
the $\Omega$ dependence of $R_2$ is significantly greater than that for 
$R_1$. To get a feeling for the level of sensitivity we note that a 
determination of $R_{1,2}$ at the level of $1-2\%$, which seems rather 
straightforward at the ILC, will be sensitive to values of $\Omega$ of order 
$-0.01$ to $-0.02$ or less. Thus even small deviations from the standard RS 
scenario should be observable at the ILC. Of course, a detailed analysis 
accounting for accelerator and detector effects should be performed to confirm 
these conclusions.

\section{Black Hole Production}

The possibility that TeV scale black holes(BH) may be a copious signal for 
extra dimensions at future colliders has been discussed by a number of 
authors{\cite {Dim,Gidd}} for both the flat and warped background cases. 
A leading approximation for the subprocess cross-section 
for the production of a BH of mass $M_{BH}$ is just the geometric 
BH size{\cite {GR,Rychkov,Kanti}} 
\begin{equation}
\hat \sigma \simeq \pi R_s^2\,,
\end{equation}
where $R_s$ is the $(4+n)$-dimensional Schwarzschild radius corresponding to 
the mass $M_{BH}$. The production of BH at the LHC has been 
studied in detail, including detector effects, in 
Refs.{\cite {Harris,Tanaka}} and there is reason to 
believe that BH arising from warped and flat backgrounds may be experimentally 
distinguishable{\cite {Stojkovic}} due to the RS $S^1/Z_2$ orbifold symmetry 
or the effects of the $AdS_5$ curvature.  
In our case of interest, we would like to examine how 
this cross section is influenced by including G-B terms in the 
RS model with $n=1$. This can be done by following, \eg, 
the analyses presented in 
{\cite {Anchor,Barrau,Cai,Myers,Rychkov2,Karasik}}. 
A simple approximate expression for the cross section can  
be obtained through a modification of the flat space result 
provided that $(R_sk\epsilon)^2 << 1$ so that the 
bulk curvature corrections can be neglected; of course we will also need to 
assume that $(\epsilon R_s/2\pi r_c)^2<<1$ so that the BH does not feel the 
finite size of the compactified space. (Recall that in the usual flat 
space case $R_s <<2\pi R_c$, where $R_c$ is the compactification radius of 
the extra dimension.) This is just the ordinary 
Schwarzschild solution in an asymptotically 
flat 5-d background{\cite {Anchor}}. We find that this approximation 
works reasonably well for small values of $\Omega$ but fails when $\Omega$ 
nears its lower limit $\Omega=-0.5$ since the $AdS$ curvature is larger 
in that case. To obtain the full expression{\cite {Cai}} these bulk 
curvature terms can no longer be neglected and 
we must consider the BH to be embedded in (an infinite!) $AdS_5$; note 
that this expression 
still assumes that $Z=(\epsilon R_s/2\pi r_c)^2<<1$, which remains as an 
approximation. We will see that this inequality is reasonably well satisfied 
below. Once we 
adjust for the definitions of the fundamental parameters as given in the 
action above, one obtains
\begin{equation}
\hat\sigma ={\pi\over {2\beta M_*^2}}\Bigg[-1+\Bigg(1-4\beta(2\alpha-
\gamma)\Bigg)^{1/2}\Bigg]\,,
\end{equation}
where 
\begin{eqnarray}
\beta &=&{k^2\over {M^2}}\Big(1-2\alpha {k^2\over {M^2}}\Big)\nonumber \\
\gamma &=&{M_{BH}\over {3\pi^2 M_*}}\,,
\end{eqnarray}
and where $M_*$ is the `warped-down' fundamental scale, $M_*=M\epsilon$. 
This expression is not very transparent so let us expand the square root to 
first order; we then find 
\begin{equation}
\hat\sigma \simeq {M_{BH}\over {3\pi M_*^3}} \Bigg[1-6\pi^2\alpha 
{M_*\over {M_{BH}}}\Bigg]\,,
\end{equation}
which is the leading order term in the bulk curvature expansion; the usual RS 
result in the flat space approximation is obtained by 
setting $\alpha=0$ in this expression{\cite {Anchor}}. 

To evaluate the cross section we note that Eq.(33) 
tells us that $\alpha$ is determined provided we know both $\Omega$ and 
$c=k/\mpl$. Note that for $k/\mpl$ fixed, $\alpha$ becomes quite large (and 
negative) as $\Omega \to -0.5$. Given the large prefactor of $6\pi^2$ in 
the approximate expression above we 
might expect that G-B terms can be numerically quite important even for small 
values of $\Omega$. It is interesting to note that this cross section is 
bounded from below as can be seen from the exact expression, \ie, 
$\hat \sigma \gsim -2\pi \alpha/M_*^2$, independent of the 
BH mass. (Remember that $\alpha$ is negative here.) In obtaining 
our numerical results we will make use of the exact expression above. 

\begin{figure}[htbp]
\centerline{
\includegraphics[width=8.5cm,angle=90]{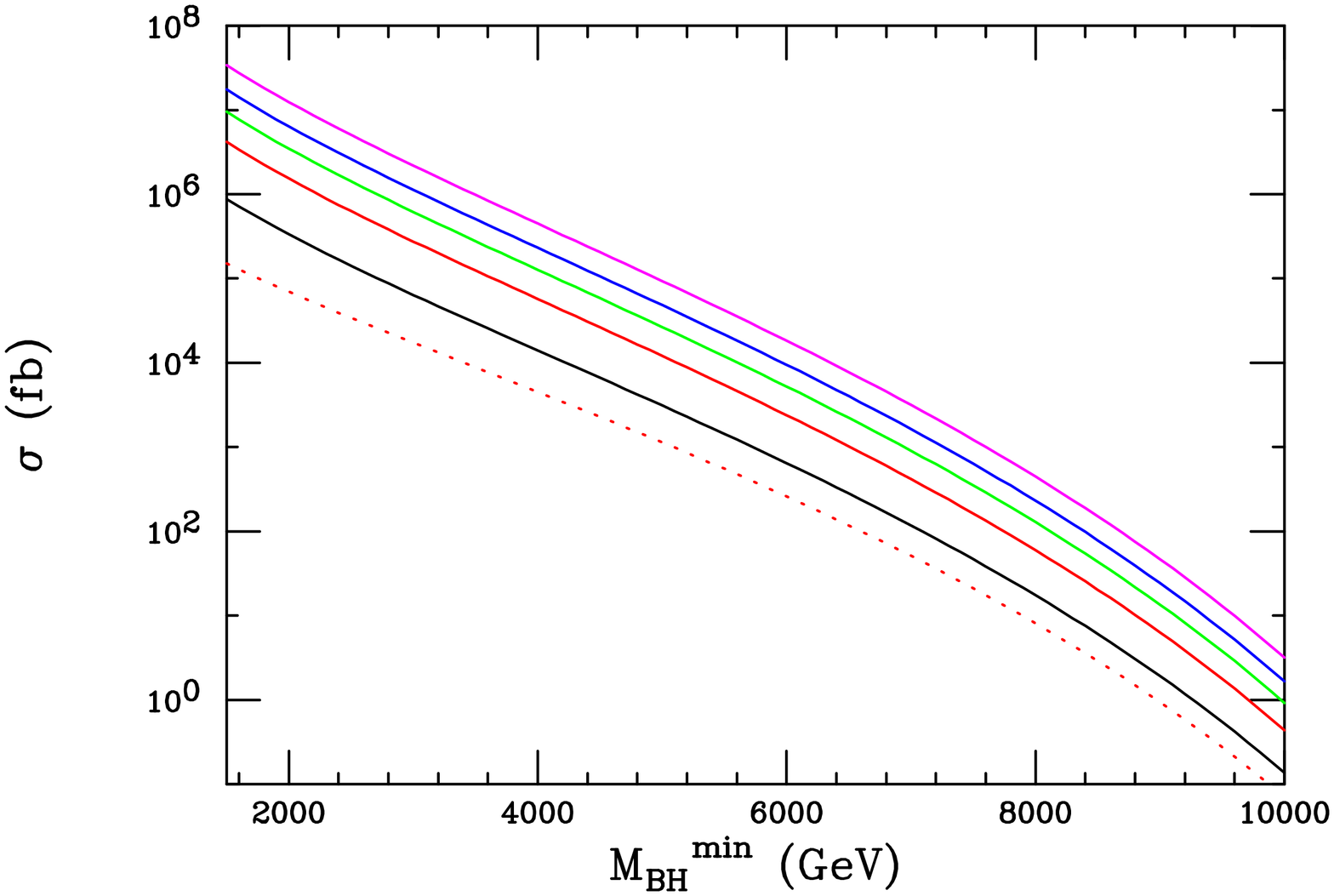}}
\vspace{0.4cm}
\centerline{
\includegraphics[width=8.5cm,angle=90]{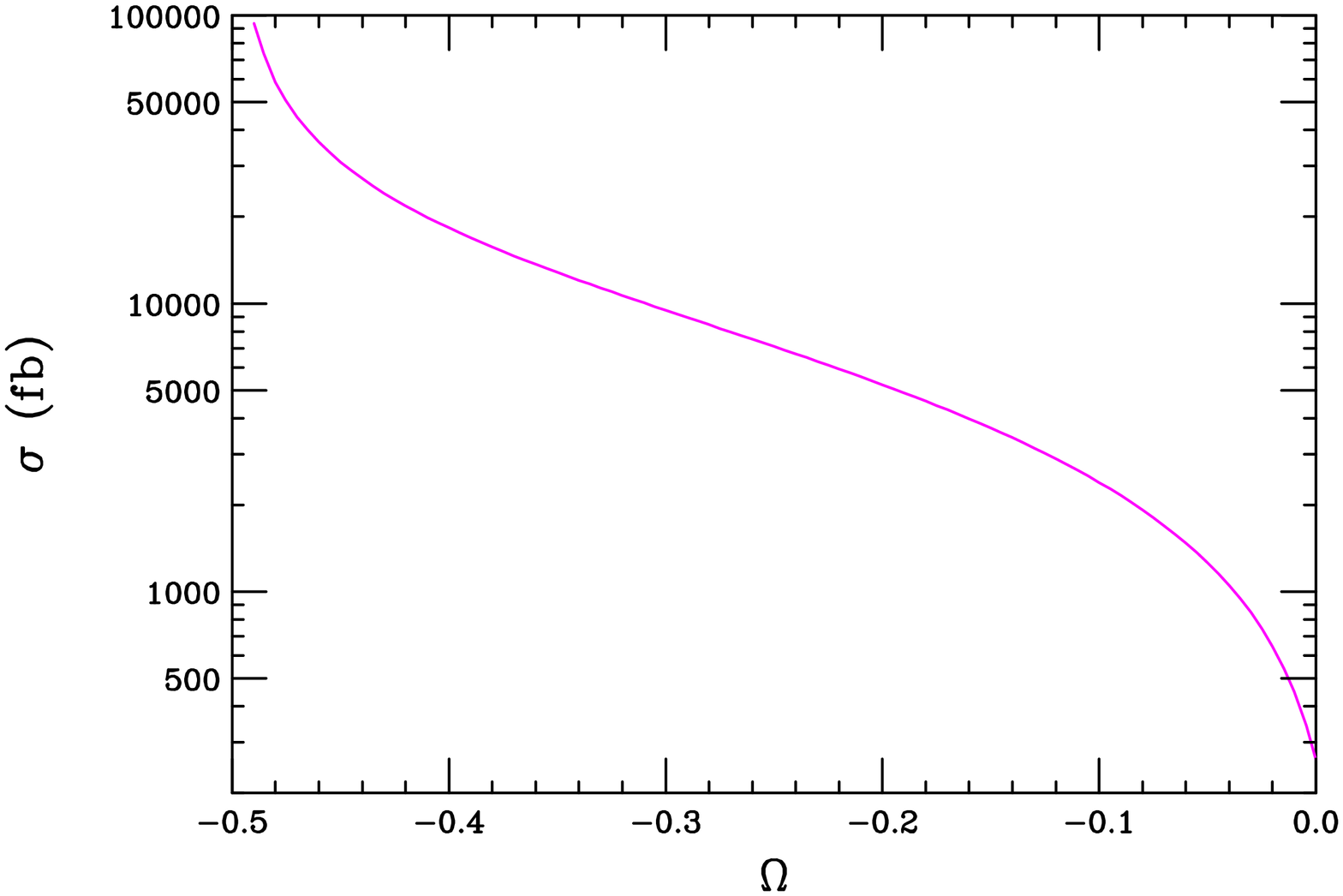}}
\vspace*{0.1cm}
\caption{(Top) Black hole 
production cross section at the LHC assuming $M_*=1.5$ TeV 
and $k/\mpl=0.1$ as described in the text. 
The lowest dotted curve corresponds to the RS model prediction with 
$\Omega=0$;  the subsequently higher 
curves corresponds to $\Omega=-0.02, -0.1, -0.2, -0.3$ and $-0.4$, 
respectively in the G-B case. (Bottom) Same as in the top panel but now for a 
fixed minimum black hole mass of 6 TeV as a function of $\Omega$.}
\label{fig8}
\end{figure}

Fig.~\ref{fig8} shows the influence of the G-B terms on the BH production 
cross-section at the LHC; for definiteness we have assumed $M_*=1.5$ TeV and 
$k/\mpl=0.1$ but there is nothing special about these particular values. We 
see, as expected, that a non-zero $\Omega$ can lead to a significant increase 
in the BH production cross section; for $\Omega=-0.1(-0.4)$ this increase is 
seen to be approximately one(two) order(s) of magnitude in comparison to the 
usual RS scenario. Here we may need to worry that the approximation that the 
effects of the finite radius of compactification can no longer be neglected.  
The inequality $Z=(\epsilon R_s/2\pi r_c)^2<<1$ is not so obviously 
satisfied here since the value of $R_s$ is growing so large. To test 
this let $\lambda$ be the ratio of the 
subprocess cross-section in the presence of G-B terms to that in the RS model. 
Then the quantity $Z$ can be expressed as 
\begin{equation}
Z=\lambda \Bigg({{M_*R_s}\over {2\pi kr_c}}\Bigg)^2 \Bigg({k^2\over {M^2}
}\Bigg)
\,,
\end{equation}
so that, with $k^2/M^2=1/20$ and $M_{BH}=5M_*$ for purposes of demonstration,  
we arrive at $Z\simeq 1.7(\lambda/100)\times 10^{-4}$ which is significant 
only for very large $\lambda$. Thus we see that for almost all of our 
parameter space $Z<<1$ remains valid. 

\begin{figure}[htbp]
\centerline{
\includegraphics[width=8.5cm,angle=90]{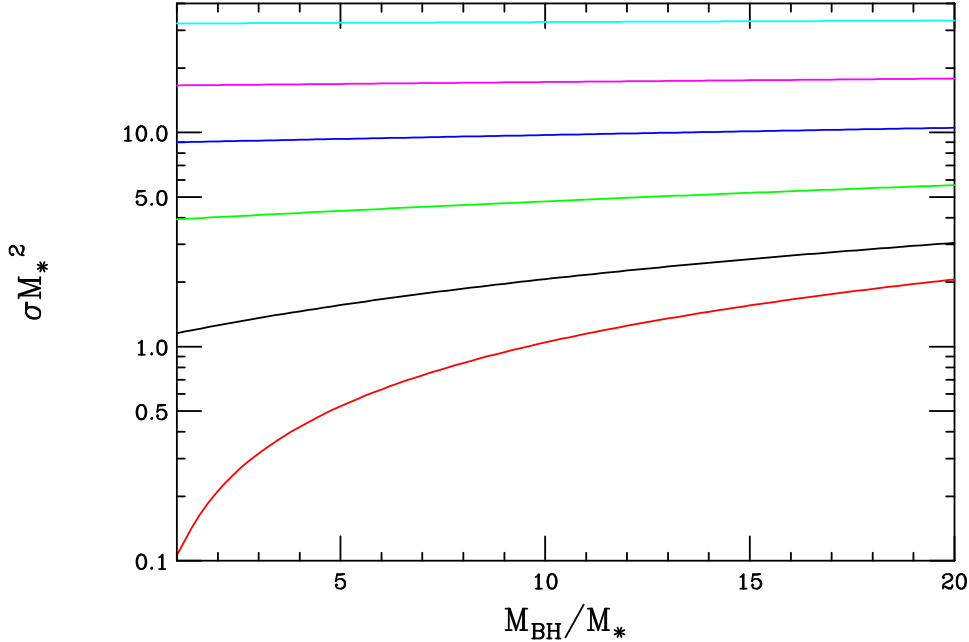}}
\vspace*{0.1cm}
\caption{The BH sub-process cross section as a function of the ratio 
$M_{BH}/M_*$ for different values of $\Omega$. From bottom to top the curves 
correspond to $\Omega=0,-0.03,-0.1,-0.2,-0.3$ and $-0.4$, respectively.}
\label{fig9}
\end{figure}

Although we have seen a substantial increase in the BH cross section relative 
to the RS model due to G-B terms, the overall normalization of the 
cross section remains uncertain due to the complexities of BH formation 
by terms of order unity as has been discussed by many 
authors{\cite {GR,Rychkov,Kanti}}. While overall factors may be uncertain 
we might expect the various scaling laws to remain at least approximately 
valid. In the case of flat extra dimensions, or in the RS model in the 
$(R_sk\epsilon)^2 << 1$ limit, the 
value of $n=d-4$ uniquely determines the simple dependence of the 
cross section 
on the ratio $M_{BH}/M_*$: $\sigma M_*^2 \sim  [M_{BH}/M_*]^{2/(n+1)}$. 
Here, for the case of  a single warped extra 
dimension, the G-B terms conspire to modify the shape of the cross section
as a function of $M_{BH}/M_*$ in a rather unique manner. This can even be 
seen by examining the approximate expression Eq.(37): the subprocess 
cross section is generally 
no longer proportional to a single power of $M_{BH}/M_*$. In fact, as the G-B 
terms come to dominate, the subprocess cross section becomes 
essentially independent of 
$M_{BH}$ as shown in Fig.~\ref{fig9} where the important feature to notice is 
the shape of the distribution and not the overall magnitude. Note 
that this effect happens quite 
rapidly as a non-zero value of $\Omega$ is turned on. This rather distinctive  
behavior may be observable in BH production at the LHC once the parton 
distribution luminosity is unfolded and will help 
to pin down the existence of G-B interactions. It should then be possible to 
correlate shifts in the production properties of BH with those of graviton 
resonances to precisely determine the G-B parameter $\Omega$.

\section{Conclusions}

Higher-dimensional curvature corrections to Einstein gravity are to be 
expected on rather 
general grounds; the observation of any effect arising from such  
terms will tell us important information about the ultraviolet completion of 
this theory. The requirements of unitarity and the absence 
of ghosts in the 5-d Randall-Sundrum model uniquely determine the form such 
terms may take and eliminates any derivatives higher than second in the 
equations of motion. Consequentially, the Gauss-Bonnet invariant is singled 
out as the only possible new non-trivial 
addition to the bulk action for gravity beyond 
the usual Ricci scalar. The existence of G-B terms in the action has been 
previously shown to leave the qualitative features of the RS model invariant   
but the quantitative details are expected to be altered, perhaps 
significantly.  

In this paper we have begun to examine the phenomenological 
implications of the existence of additional G-B terms in the RS bulk action. 
Two of the dominant signatures for the RS model are the production of 
TeV-scale graviton resonances and the production of 
black holes at future colliders. This paper 
has shown that both of these processes can be significantly altered if G-B 
terms are present in any appreciable amount. 
If no deviations from the classic RS model predictions are observed, 
we have shown that stringent bounds can be placed on the parameter 
$\Omega$ which describes the relative strength of the G-B terms.  

The first step in this analysis was to determine how the 
graviton KK masses, couplings and wavefunctions are modified by the presence 
of the G-B terms. Once this was done we analyzed the production properties 
of the KK graviton resonances at the ILC; we showed the mass spectrum and 
decay widths of these resonances were substantially different than in the RS 
case over most of the G-B model parameter space. It was demonstrated that 
the KK mass and width ratio 
measurements at the ILC can be highly sensitive to the presence of G-B 
terms and can be used to precisely determine their strength or to 
place strong bounds on their 
existence if no deviations from the RS model are observed. 

Next we examined the production of BH at the LHC in the RS model with G-B 
terms. We found that not only is the production cross section significantly 
enhanced in comparison to the conventional RS model but its parametric 
dependence on the ratio $M_{BH}/M_*$  can be drastically altered even for 
small values of $\Omega$. Any deviations observed in the production of 
graviton resonances at the LHC or ILC 
can then be directly correlated with modifications to BH production to 
determine the value of $\Omega$.

Hopefully signals of warped extra dimensions will be observed at future 
colliders and we may be ultimately be 
able to probe the ultraviolet completion of quantum gravity.

\noindent{\Large\bf Acknowledgments}

The author would like to thank J.Hewett and B. Lillie for discussions 
related to this work.

\section*{Appendix}

In this Appendix we consider a modification of the action given by Eqs.(3) 
and (4) above to include graviton 
brane kinetic terms generated by the 4-d Ricci 
curvature scalars on the TeV and Planck branes. Such terms have been suggested 
in the case of G-B extensions of the RS model{\cite {Brax}} to help assist 
removal of the tachyon field present{\cite {Charmousis}} when $\alpha>0$.  
Using the notation above 
as well as in Ref.{\cite {DHRbt}}, this amounts to adding to the action a term 
of the form
\begin{equation}
S_{BKT}=\int d^5x \sqrt {-g} ~{M^3\over {2}} \Big[g_0r_c \delta(y)
+g_\pi r_c\delta(y-\pi r_c)\Big]R^{(4)}\,,
\end{equation}
where $R^{(4)}$ is the induced 4-d curvature and $g_{0,\pi}$ are dimensionless 
constants that we expect to be of order unity. Since these are 
boundary terms they do not modify the bulk equation for the KK wavefunctions 
which remain unaltered from the traditional RS case as discussed above. 
Defining the combinations 
\begin{equation}
\gamma_{0,\pi}=g_{0,\pi}{{kr_c}\over {2}}\,,
\end{equation}
as in our earlier work{\cite {DHRbt}}, we find that the 
only change induced by these new 
brane terms is to modify the definition of $w(y)$ given in Eq.(18) above to 
\begin{equation}
w(y)=e^{-2\sigma}\Bigg(1-4\alpha {k^2\over {M^2}}+8\alpha {k\over {M^2}}
\Delta+{{2\gamma_0}\over {k}}\delta(y)+{{2\gamma_\pi}\over {k}}
\delta(y-\pi r_c)\Bigg)\,, 
\end{equation}
with the subsequent appropriate changes in the boundary conditions: 
\begin{equation}
\psi_n(0^+)~'+\Omega_0{m_n^2\over {k}}\psi(0^+)=0\,,
\end{equation}
and 
\begin{equation}
\psi_n(\pi r_c^+)~'+\Omega_\pi \epsilon^{-2}{m_n^2\over {k}}
\psi_n(\pi r_c^+)=0\,,
\end{equation}
where now we now define the combinations
\begin{equation}
\Omega_{0,\pi}={{4\alpha k^2/M^2\pm\gamma_{0,\pi}}\over {1-4\alpha k^2/M^2}}\,.
\end{equation}
The generalizations of the other equations in the main text can now be read 
off directly, \eg, 
\begin{equation}
\zeta_1(x_n)+\Omega_\pi x_n \zeta_2(x_n)=0\,,
\end{equation}
and 
\begin{equation}
\beta_n=-{{J_1(x_n\epsilon)+\Omega_0 x_n\epsilon J_2(x_n\epsilon)}\over 
{Y_1(x_n\epsilon)+\Omega_0 x_n\epsilon Y_2(x_n\epsilon)}}\,, 
\end{equation}
determine the masses and wavefunctions of the KK states. 
A short analysis shows that there will exist parameter ranges that are tachyon 
free.  The collider and black hole analyses will now require a detailed 
exploration  of this $\Omega_0-\Omega_\pi$ parameter space which is beyond 
the scope of the present paper.

%
%%%%%%%%%%%%%%%%%%--- References
%%%%%%%%%%%%%%%%%%%%%%%%%%%%%%%%%%%%%%%%%%%%%%%%%%%%%%%
\def\MPL #1 #2 #3 {Mod. Phys. Lett. {\bf#1},\ #2 (#3)}
\def\NPB #1 #2 #3 {Nucl. Phys. {\bf#1},\ #2 (#3)}
\def\PLB #1 #2 #3 {Phys. Lett. {\bf#1},\ #2 (#3)}
\def\PR #1 #2 #3 {Phys. Rep. {\bf#1},\ #2 (#3)}
\def\PRD #1 #2 #3 {Phys. Rev. {\bf#1},\ #2 (#3)}
\def\PRL #1 #2 #3 {Phys. Rev. Lett. {\bf#1},\ #2 (#3)}
\def\RMP #1 #2 #3 {Rev. Mod. Phys. {\bf#1},\ #2 (#3)}
\def\NIM #1 #2 #3 {Nuc. Inst. Meth. {\bf#1},\ #2 (#3)}
\def\ZPC #1 #2 #3 {Z. Phys. {\bf#1},\ #2 (#3)}
\def\EJPC #1 #2 #3 {E. Phys. J. {\bf#1},\ #2 (#3)}
\def\IJMP #1 #2 #3 {Int. J. Mod. Phys. {\bf#1},\ #2 (#3)}
\def\JHEP #1 #2 #3 {J. High En. Phys. {\bf#1},\ #2 (#3)}

\end{document}